\documentclass[final,5p,times,twocolumn]{elsarticle}

\usepackage[utf8x]{inputenc}
\usepackage{amssymb}
\usepackage{amsmath}
\usepackage{amsthm}
\usepackage{amsmath}
\usepackage{hyperref}
\usepackage{subfigure}
\usepackage[percent]{overpic} 
\usepackage{picture}
\usepackage{color}
\usepackage{nth}
\usepackage{gensymb}
\usepackage{lineno} 
\usepackage{textcomp}
\usepackage{threeparttable}

\graphicspath{{img/}}

\usepackage[figuresright]{rotating}

\journal{Nuclear Instruments and Methods in Physics Research A }

\begin{document}

\begin{frontmatter}


\title{Overview of Plasma Lens Experiments and Recent Results at SPARC$\_$LAB}


\author[lnf]{E. Chiadroni}\ead{enrica.chiadroni@lnf.infn.it}
\author[lnf]{M.P. Anania}
\author[lnf]{M. Bellaveglia}
\author[lnf]{A. Biagioni}
\author[lnf]{F. Bisesto}
\author[lnf]{E. Brentegani}
\author[lnf]{F. Cardelli}
\author[roma2]{A. Cianchi}
\author[lnf]{G. Costa}
\author[lnf]{D. Di Giovenale}
\author[lnf]{G. Di Pirro}
\author[lnf]{M. Ferrario}
\author[lnf]{F. Filippi}
\author[lnf]{A. Gallo}
\author[lnf]{A. Giribono}

\author[lnf]{A. Marocchino}

\author[roma1]{A. Mostacci}
\author[lnf]{L. Piersanti}
\author[lnf]{R. Pompili}
\author[ucla]{J. B. Rosenzweig}
\author[mi]{A. R. Rossi}
\author[lnf]{J. Scifo}
\author[lnf]{V. Shpakov}
\author[lnf]{C. Vaccarezza}
\author[lnf]{F. Villa}
\author[racah]{A. Zigler}

\address[lnf]{INFN-Laboratori Nazionali di Frascati, Via Enrico Fermi 40, 00044 Frascati (Rome), Italy}
\address[roma2]{University of Rome "Tor Vergata" and INFN-Roma Tor Vergata, Via della Ricerca Scientifica 1, 00133 Rome, Italy}
\address[roma1]{University of Rome "Sapienza", Piazzale Aldo Moro 5, 00185 Rome, Italy}
\address[ucla]{UCLA, Los Angeles, California 90095, USA}
\address[mi]{INFN-Milano and University of Milan, Via Celoria, I-16-20133 Milano, Italy}
\address[racah]{Racah Institute of Physics, Hebrew University, 91904 Jerusalem, Israel}

\begin{abstract}
Beam injection and extraction from a plasma module is still one of the crucial aspects to solve in order to produce high quality electron beams with a plasma accelerator. Proper matching conditions require to focus the incoming high brightness beam down to few microns size and to capture a high divergent beam at the exit without loss of beam quality. Plasma-based lenses have proven to provide focusing gradients of the order of kT/m with radially symmetric focusing thus promising compact and affordable alternative to permanent magnets in the design of transport lines. In this paper an overview of recent experiments and future perspectives of plasma lenses is reported.
\end{abstract}


\end{frontmatter}


\section{Introduction}\label{intro}
Multi-GeV acceleration, both laser and particle beam driven, has been already demonstrated in cm-scale plasma structures~\cite{geddes2004high,leemans2006gev,2007Blumenfeld,2008PhRvL.100g4802K}. Great efforts are currently ongoing in several groups~\cite{EuPRAXIApaper,ferrario2013sparc_lab,FLASHForward,BELLA} worldwide for the acceleration of high brightness electron beams, which need to be captured and 
transported up to the final application. The aim is the preservation of the quality of the 6D phase space, which is demanding in particular for FEL experiments. In this regard, both in case of plasma and RF injectors, the control of electron injection into the plasma module is mandatory for efficient acceleration: the beam must satisfy the transverse matching condition at the plasma entrance to prevent envelope oscillations that may cause emittance growth. In particular, the following condition for the Twiss parameters of the witness beam holds in the blow-out regime and assuming negligible beam loading~\cite{beamloading}):
\begin{eqnarray}
\beta_{matching}=&\frac{\sqrt{2\gamma}}{k_p}~,
\end{eqnarray}
where $k_p=2\pi/\lambda_p$ is the inverse plasma skin depth, with the plasma wavelength depending on the plasma background density as $\lambda_p(\mu m) \approx 3.3\cdot10^{10} n_{p}^{-1/2}(cm^{-3})$; $\gamma$ is the Lorentz factor for the electron beam. The matching condition for the beam transverse size is 
\begin{equation}
\sigma_{matching}=\sqrt{\frac{\beta_{matching}{\varepsilon_{n}}}{\gamma}}~:
\end{equation}
with the typical numbers involved, e.g. $\gamma$=1000, $n_p$=10$^{16}~cm^{-3}$ and $\varepsilon_n$=1~mm~mrad, the beam transverse size is of micrometer scale.

One option to relax the tight matching condition on the transverse plane is to use plasma ramps, the progressive transverse force focuses the beam along the density profile. A density shaping can be achieved by either varying the capillary diameter along its length~\cite{FilippiTapering2017}, or tailoring the gas profile~\cite{Schaper2014}, or eventually exploiting optical-ionization methods~\cite{Wittig2015}. Advantages of density ramps are discussed in~\cite{Floettmann,TomassiniRossi2016,deLaOssa2017}, however ramps are still an open question and consequently under investigation due to their cumbersome nature. We notice that ramps can be formed by non-fully ionized gas eventually non-even in thermal equilibrium, conditions that might affect bunch quality at some degree. The key difficulty that still holds is whether the ramp can have an arbitrarily long shape where the bunch would self focus and reach a natural equilibrium, or if the ramp need to have a length function of the bunch betatron function~\cite{1994PhRvE..49.4407B,MarocchinoThisConf}.

Once accelerated in the plasma, electron beams must be captured and transported along the beam line. When exiting the plasma region, electrons move from an extremely intense focusing field, generated inside the bubble, to a free space where the focusing effect suddenly vanishes. Indeed, plasma fields are stronger, $10^{2}-10^{3}$ times, than in conventional accelerators, depending on the plasma density $n_p$ as~\cite{su1990plasma}  
\begin{equation}\label{Eq:plasmagrad}
G(MT/m)\approx 3 ~n_p (10^{17}cm^{-3})~.
\end{equation}
With the typical plasma densities considered in these experiments, i.e. of the order of $10^{16}-10^{17}$~cm$^{-3}$, $G\approx 1~$MT/m. Therefore, because of mrad-scale angular divergence, the beam experiences a huge transverse size variation when propagating from the plasma outer surface to the following beam line element.
Under these conditions, the particle transverse motion becomes extremely sensitive to the energy spread: the betatron frequency of a particle critically depends on its energy, therefore particles with different energies, in a drift, rotate with different velocities in the transverse phase space, resulting in a wider bunch trace space area. As a consequence, the resulting projected normalized emittance becomes a function both of drift length and energy spread. In this regard, the beam angular divergence has to be reduced and the transverse spot size increased to limit the chromatic induced emittance degradation in free space~\cite{Merling2012,Floettmann} as formulated by~\cite{Migliorati2013}
\begin{equation}
\varepsilon_{n}^{2}=<\gamma>^2(\sigma_{E}^2\sigma_{x}^2\sigma_{x'}^2+\varepsilon^2)\approx<\gamma>^2(\sigma_{E}^2\sigma_{x'}^4 s^2+\varepsilon^2)~,
\end{equation}
where the bunch size dependence on the angular divergence in a drift is $\sigma_x(s)=\sigma_{x'}s$, assuming a beam waist as starting condition. 

For the reasons above mentioned, a radially symmetric focusing gradient of the order of kT/m, and eventually higher, is needed at the injection of the plasma accelerating module to guarantee the matching condition. In addition, it is preferable at the extraction from it to limit the emittance growth in the transition from the plasma module and the free space. Furthermore, since chromaticity scales inversely with the betatron function, any capture and transport system should be placed in the immediate proximity of the plasma outer surface to avoid beam quality degradation, if the energy spread is not mitigated below 1\%. Additional features of such a focusing device should be a focusing strength scaling proportionally to $1/\gamma$, to be effective even at ultra-relativistic energies, and a focusing field varying linearly with the radius, to prevent emittance degradation due to geometric aberrations. Finally, a focusing field independent on the beam distribution and the tunability of the focusing system to adjust the focal length would be preferred.

\section{Conventional focusing systems}\label{MagneticLens}
One option would be using conventional focusing systems, such as solenoid and permanent magnet quadrupoles (PMQ); however, they both have disadvantages, which make their use not advisable. In particular, solenoid magnets, whose advantages are the radial symmetric focusing and the tunability, are characterized by a focusing strength, which scales as $1/\gamma^2$:
\begin{equation}
K_{sol}=\Big(\frac{e_0B}{2m_0c}\Big)^2\frac{1}{\gamma^2}~,
\end{equation}
resulting in a significant contribution to chromaticity as long as the energy spread is not negligible, due to the different focusing experienced by particles in the bunch.
On the other hand, PMQs whose focusing strength scales as $1/\gamma$, with focusing gradients up to nearly 600 T/m~\cite{becker2009characterization,nichols2014analysis}, provide radial focusing only in triplet configuration, resulting in a longer, few tens of cm, effective focal length with increased chromaticity. 
State-of-the-art quadrupoles are not strong enough compared to transverse plasma gradients (see Eq.3) to limit the emittance growth that might occur because of the abrupt transition between plasma and vacuum. At the exit of the plasma-accelerating module, the beam size is a few micrometers, therefore with typical values for the normalized emittances, i.e. 1 mm mrad, the angular divergence is several mrad and the betatron function few millimeters for GeV beams. Therefore, in the drift downstream from the plasma-accelerating module, the betatron function, which increases with the square of the drift length, reaches values of the order of 10 m after 100 mm. In addition, since quadrupoles are defocusing in one of both planes, the betatron function reaches large values (~10-100 m) at the next quadrupole, resulting in chromatic effects emphasized by the large quadrupole gradient. These chromatic effects generate emittance growth and beam degradation. In addition, non-trivial movable holders are needed to adjust the focal length~\cite{lim2005adjustable}, resulting in a narrow range of beam energy covered by a single system. 

Therefore the natural choice would be the exploitation of plasma fields to focus beams.

\section{Plasma lenses}
Plasma focusing can occur because of the following mechanisms: (i) by the self-focusing due to the shielding process produced when the background plasma reorganizes itself to conserve the overall neutrality after the passage of a driver beam (passive plasma lens), and (ii) by the azimuthal magnetic field produced by an externally-driven axial current (active plasma lens).

In 1922 electrostatic focusing of a continuous low energy electron beam was observed due to electrostatic fields created by a beam-ionized gas within a cathode ray tube~\cite{JBJohnson}. In the early 1930s the concept of passive plasma lenses was conceived, showing that an electron stream can magnetically self-focus, if it has sufficient current and its space charge is neutralized by positive ions~\cite{Bennett}. 
Two regimes can be exploited, defined by the ratio between the beam density, $n_b=\frac{N}{(2\pi)^{3/2}\sigma_z\sigma_{r}^{2}}$, and the plasma density, $n_p$ ($N$ is the number of particles in the bunch, $\sigma_z$ and $\sigma_r$ the beam longitudinal and transverse size, respectively). In the linear or over-dense regime, $\frac{n_{b}}{n_p}\ll 1$, the bunch's self-field creates only a small perturbation of the plasma density~\cite{rosenzweig1991acceleration}: the plasma electrons respond to the excess of charge by moving away from the beam particles, while the remaining plasma ions neutralize the space charge force within the beam, resulting in the beam focusing due to its self-generated azimuthal magnetic field, whose focusing strength depends only on the beam density as\begin{equation}
K=\frac{2\pi r_{e}n_b}{\gamma}~.
\end{equation}
In the blow-out or under-dense regime, $\frac{n_b}{n_p} \gg 1$, the plasma response is not well described by the linear fluid theory; the plasma electrons are completely rarefied by the electron beam, which drives a strong plasma wave in the background plasma; the electron beam is focused by the uniform charge density of the ions, resulting in a linear focusing, nearly aberration-free, due to the transverse electric fields of the plasma wave, and the focusing strength depends only on the plasma density and not on the local beam density~\cite{barov1994propagation}:
\begin{equation}
K=\frac{2\pi r_{e}n_p}{\gamma}~.
\end{equation}

In mid-1980s, P. Chen first proposed passive plasma lenses as final focus elements~\cite{PChen87} to improve the luminosity in future high energy e+e- colliders because of the ultra-strong, up to MT/m, field gradient. Passive plasma lenses, being the focusing force governed by the densities of both, the beam and the plasma, are characterized by a limited tunability, with beam parameters affecting the focusing force and the lens aberrations. 



The first idea of using externally driven plasma axial current to focus a proton beam was conceived by Panofsky and Baker in 1950~\cite{panofsky1950focusing}. Focusing is due to plasma discharge current either in gas-filled capillary or in an evacuated channel by the azimuthal magnetic field. Because of the external discharge current, this plasma lens has been named as active. 

An active plasma lens behaves as a current carrying conductor, realized by means of a discharge applied between the electrodes at the edges of a capillary. The bunch is focused by the azimuthal magnetic field, $B_{\phi}$, generated by the discharge current according to Ampere's law:
\begin{equation}
B_{\phi}(r)=\frac{\mu_0}{r}\int_{0}^{r}J(r')r'dr'~,
\end{equation}
where $\mu_0$ is the vacuum permeability and $J(r)$ the current density within the aperture ($r<R$, with $R$ is the capillary radius). The focusing strength is then given by 
\begin{equation}
K=\frac{\partial B_{\phi}(r)}{\partial r}\frac{e_0}{m_0 c \gamma}~,
\end{equation}
and, in principle, optimal focusing condition is reached when the current density is perfectly parallel to the capillary axis and transversely uniform: in this case, the magnetic field intensity has a linear dependence on the distance from the axis. Active plasma lenses have tunable focusing strength, being dependent on the applied discharge current, which can be changed by delaying the beam arrival time with respect to the beginning of the discharge. 

\begin{figure}[h]
\centering
\includegraphics[width=0.7\linewidth]{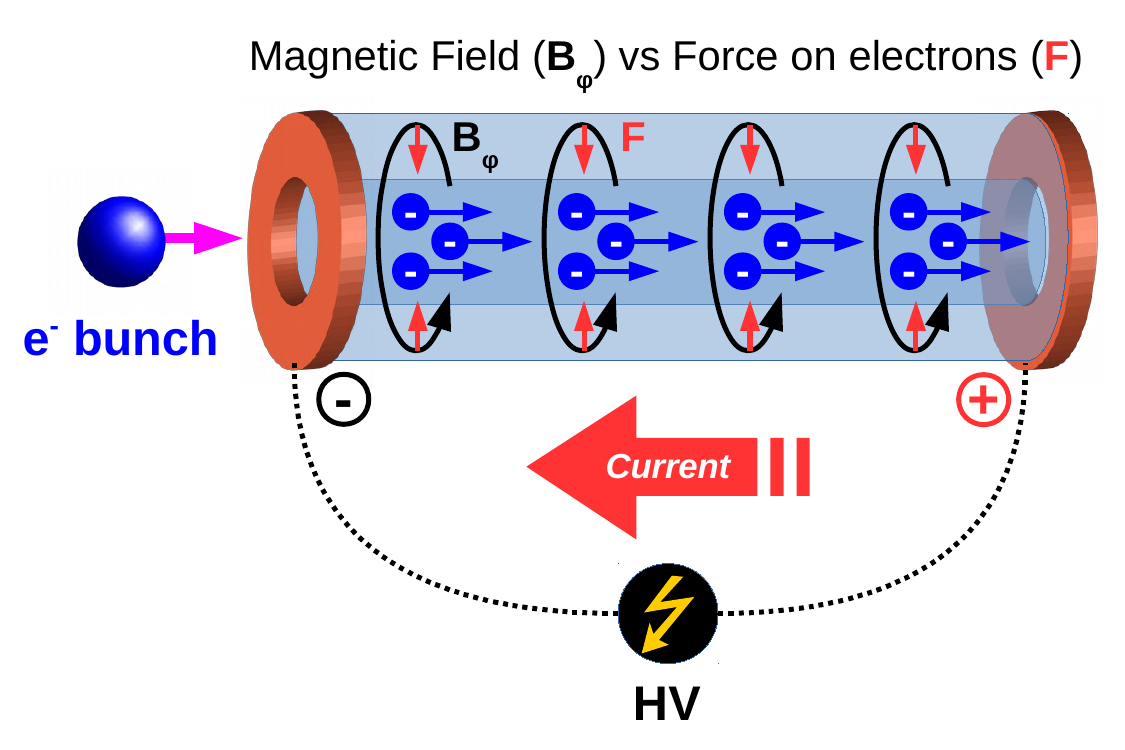}
\caption{Schematic picture of the active plasma lens and its focusing feature.}
\label{Fig:APL}
\end{figure}

A schematic picture of an active plasma lens is depicted in Fig.~\ref{Fig:APL}.

Since beam injection and extraction from a plasma accelerating module is still one of the crucial aspects to be solved for the plasma-based acceleration of high quality electron beams, plasma lenses are currently enjoying a renaissance worldwide to demonstrate their integration in conventional transport beam lines without affecting the beam quality and, in particular, avoiding emittance degradation. 

In 2010s the BELLA group at LBNL first reassessed the idea of active plasma lenses~\cite{van2015active} for staging of jet-based laser plasma accelerators (LPA)~\cite{Barber2017}, demonstrating the symmetric focusing with gradient of the order of kT/m and highlighting the effect on the beam transverse distribution of non-uniformity of the discharge current~\cite{van2017active}. In the same period at the SPARC$\_$LAB test facility~\cite{ferrario2013sparc_lab}, we started investigating active plasma lenses paying attention to the emittance characterization downstream from the plasma~\cite{pompili2017experimental} and highlighting the passive lens focusing effect experienced by the electron beam at delay times (from the beginning of the discharge) when the current is too low to produce any active focusing, and the self-focusing gradient dominates due to the over-dense regime~\cite{lens_alberto}.
Recently a FLASHForward collaboration has conducted experiments at the Mainz Microton (MaMi) with 855 MeV beam, providing the first direct measurement of the magnetic field gradient~\cite{Roeckemann2017}. More recently experiments are ongoing at the CLEAR~\cite{CLEAR} user facility at CERN with the aim of studying the presence of plasma wakefields and how they eventually distort the beam distribution. 

In the following section experiments performed at the SPARC$\_$LAB test facility will be presented with the aim of providing a full comprehension of the focusing process, which depends on several aspects, e.g. plasma temperature, ionization degree, that contribute to a more transversely non-linear magnetic field profile. The only way to demonstrate the linearity of the magnetic field within the capillary radius is given by the measurement of the normalized projected emittance downstream from the plasma which, in principle, does not degrade beam parameters, being the scattering in the plasma negligible. 

\section{Experimental Results at SPARC$\_$LAB}

SPARC$\_$LAB is a test facility based on a high brightness photo-injector and a high power laser (200 TW, $<$30 fs pulse) to drive Thomson back-scattering~\cite{vaccarezza2016sparc_lab} and THz radiation~\cite{giorgianni2016tailoring} sources, Free-Electron Laser (FEL) experiments~\cite{Ronsivalle2013} and plasma-based acceleration experiments, both laser-driven~\cite{rossiplasma} and particle-driven~\cite{ChiadroniCuba2017}. The updated photo-injector layout consists of two S-band traveling wave (TW) accelerating structures and one TW, constant impedance, C-band structure~\cite{AlesiniCband}, which for the plasma lens experiments shown here has been kept off, providing only a drift and no acceleration. The plasma interaction chamber is fully equipped with diagnostics, both transverse and longitudinal, with a $H_2$ plasma discharge capillary\footnote{The H$_2$ generation and injection system consists of an Electrolytic generator, a pressure reduction system (300 mbar become 10 mbar in capillary), an electro-valve triggered by the HV discharge with tunable aperture (3 ms), to let gas flow inside the capillary, and delay time (10 us before discharge). }~\cite{anania2016plasma} and permanent magnet quadrupoles for beam matching in and out from the plasma for plasma acceleration experiments. The photo-injector can be set, depending on the experiment, to provide energy between 30 and 160 MeV, with bunch charge ranging between ten and multi-hundreds of pC in fs up to ps bunch duration. For the plasma lens experiment presented here the achieved beam parameters are listed in Table~\ref{tab:Beam}.

At SPARC$\_$LAB we have started investigating active plasma lenses with the aim of characterizing the effect on the transverse emittance, which is of utmost importance if plasma lenses need to be integrated in conventional transport beam lines. In the previous work~\cite{pompili2017experimental}, we have observed a dramatic increase of the trasverse emittance probably due to the fact that the magnetic field profile is linear only close to the axis, as it is also confirmed by simulations. To describe the main effects of the discharge process, we have followed a one-dimensional analytical model~\cite{bobrova2001simulations} that assumes the distribution of plasma inside the capillary at the equilibrium stage as soon as the discharge is initiated. 
The magnetic field profile assumes that the equilibrium is determined only by the balance between Ohmic heating and cooling due to the electron heat conduction. With the 3~cm long capillary and the lower peak current (20 kV, 100 A) discharge circuit, we have observed, as reported in~\cite{pompili2017experimental}, a partial ionization of the gas, being maximum of 30\% in close proximity to the axis. The partial ionization of the hydrogen results in an even more non-linear magnetic field, since the current is forced closer to the axes. To get rid of the non-linearities in the magnetic field, causing emittance degradation, we replaced the 3 cm long capillary with a shorter one, 1 cm length\footnote{The 1 cm-long capillary, with 1~mm hole diameter, is made by 3D printing; one inlet is open for gas flow at 1/2 of capillary.}, and we modified the discharge circuit to produce a peak discharge current of 240 A with 20 kV, resulting in an unchanged focal length of 20 cm, where a YAG:Ce screen is placed for transverse beam size measurement. 

\begin{figure}[h]
\centering
\includegraphics[width=0.9\linewidth]{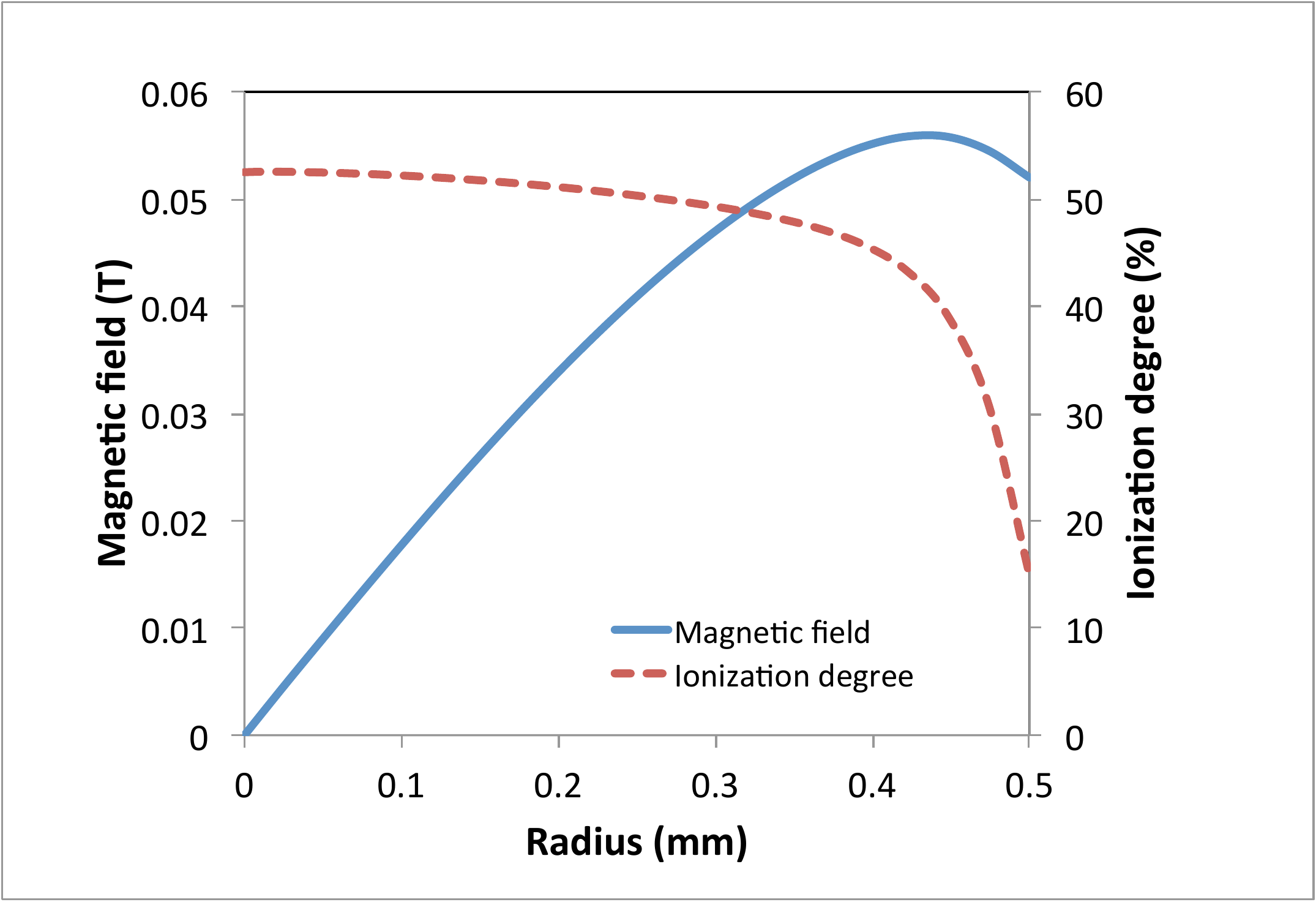}
\caption{Calculated radial profiles of the azimuthal magnetic field (blue) and $H_2$ ionization degree (red) for 130 A discharge current, value at which the beam transverse size on the
screen is minimized.}
\label{Fig:IonizDegree}
\end{figure}

The one-dimensional model used confirms a higher and more uniform ionization degree within the capillary radius as depicted in Fig.~\ref{Fig:IonizDegree} (red curve). The current is distributed approximately uniformly within the capillary aperture, resulting in a more linear magnetic field (Fig.~\ref{Fig:IonizDegree} (blue curve)).

\begin{figure}[h]
\centering
\includegraphics[width=0.9\linewidth]{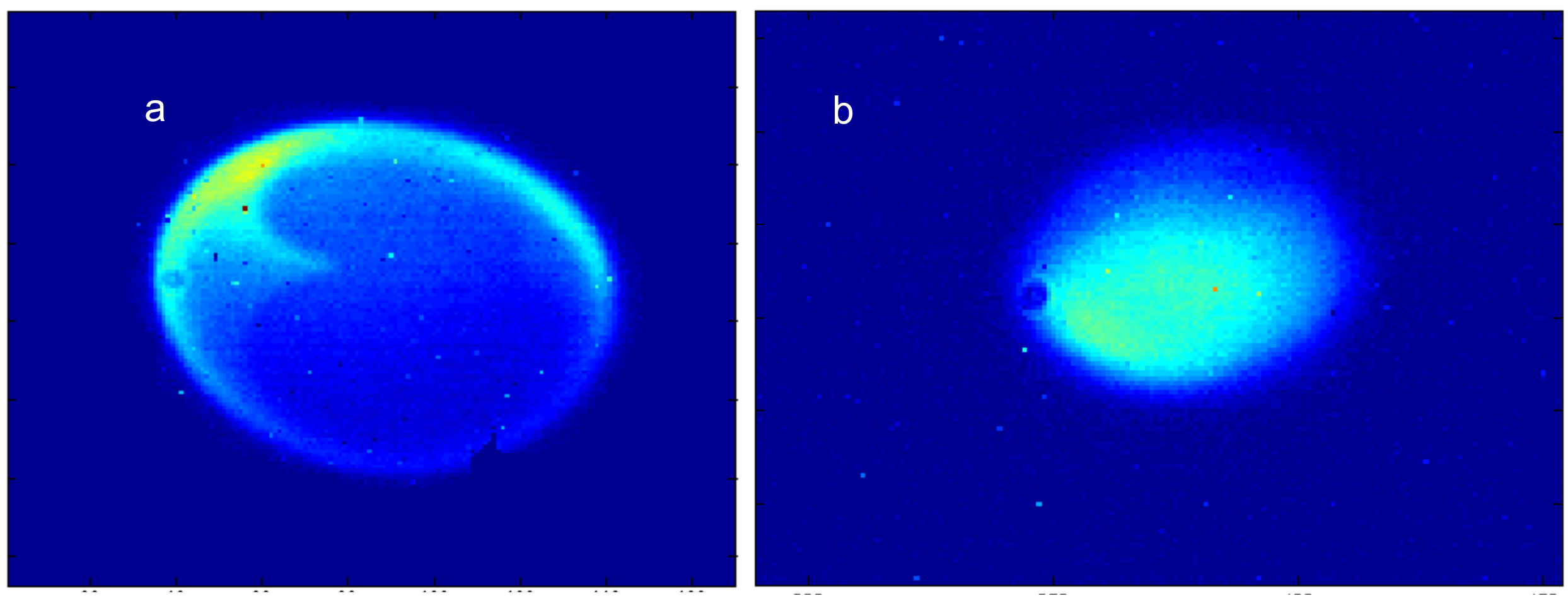}
\caption{Beam transverse distribution measured on the YAG screen at 20~cm from the capillary for the maximum current: (a) 100 A and 3 cm length, (b) 240 A and 1 cm length.}
\label{Fig:spotcomp}
\end{figure}

Under these conditions, and with beam parameters at the plasma entrance listed in Table~\ref{tab:Beam}, we have observed a more uniform beam transverse distribution as shown in Fig.~\ref{Fig:spotcomp} for the two cases. In particular, at the current of maximum defocalization, the beam distribution is not affected by spherical aberrations (Fig.~\ref{Fig:spotcomp}b), signature of non-linear magnetic field and responsible of the emittance growth observed in~\cite{pompili2017experimental}. 
\begin{table}[hbt]
\center{
\caption{Measured beam parameters at the plasma entrance.}
  \label{tab:Beam}
\begin{tabular}{lc}
\hline
Q (pC) & 50 (5)\\
E (MeV) & 126.5 (0.04)\\ 
$\Delta E/E $ (\%) & 0.06\\
$\varepsilon_{nx}$ (mm~mrad) & 0.9 (0.1)\\
$\varepsilon_{ny}$ (mm~mrad) & 1.15 (0.05)\\
$\sigma_z$ ($\mu$m) & 303 (6)\\
$\sigma_{x}$ ($\mu$m) & 79 (2)\\
$\sigma_{y}$ ($\mu$m) & 86 (2)\\
\hline
\end{tabular}
}
\end{table}
Figure~\ref{Fig:spotcomp} shows a comparison between measured transverse distributions in case of an ionization degree of about 30\% mainly on axis (a) and of 50\% over a larger fraction of the capillary radius (b).

\begin{figure}[h]
\centering
\includegraphics[angle=-90,width=0.9\linewidth]{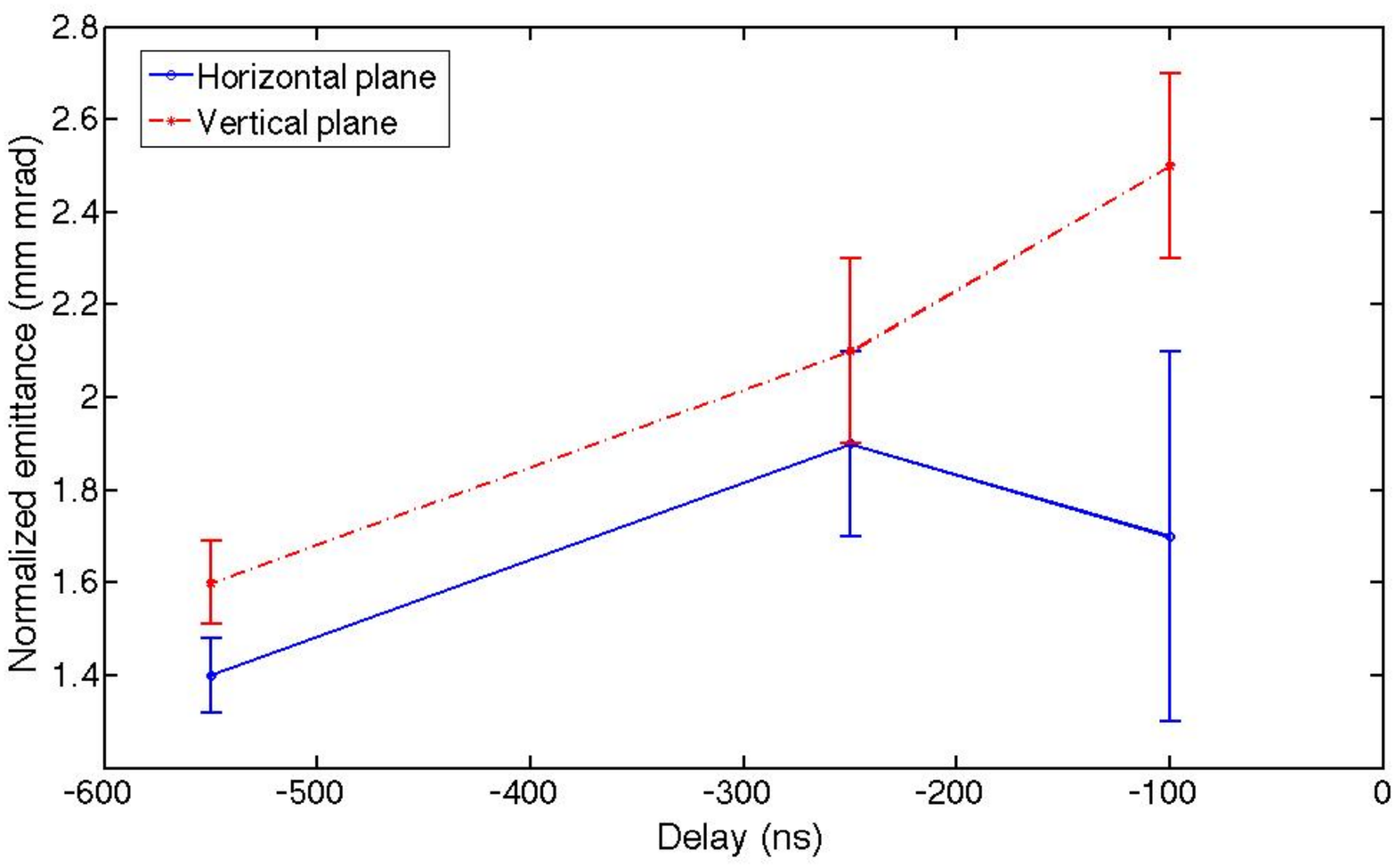}
\caption{Measured normalized transverse emittance as function of the time delay from the beginning of the discharge.}
\label{Fig:emittance}
\end{figure}

The more linear magnetic field within the capillary contributes to a moderate emittance growth as experimentally observed. The emittance, measured as function of the time delay from the beginning of the discharge, is reported in Fig.~\ref{Fig:emittance}. The arrival time of the electron beam is scanned with respect to the discharge pulse in order to change the active plasma lens focusing. The -550 ns delay corresponds to the current value that produces a beam waist at the screen. 

\begin{figure}[h]
\centering
\includegraphics[bb=150 100 370 500,angle=-90,width=0.85\linewidth]{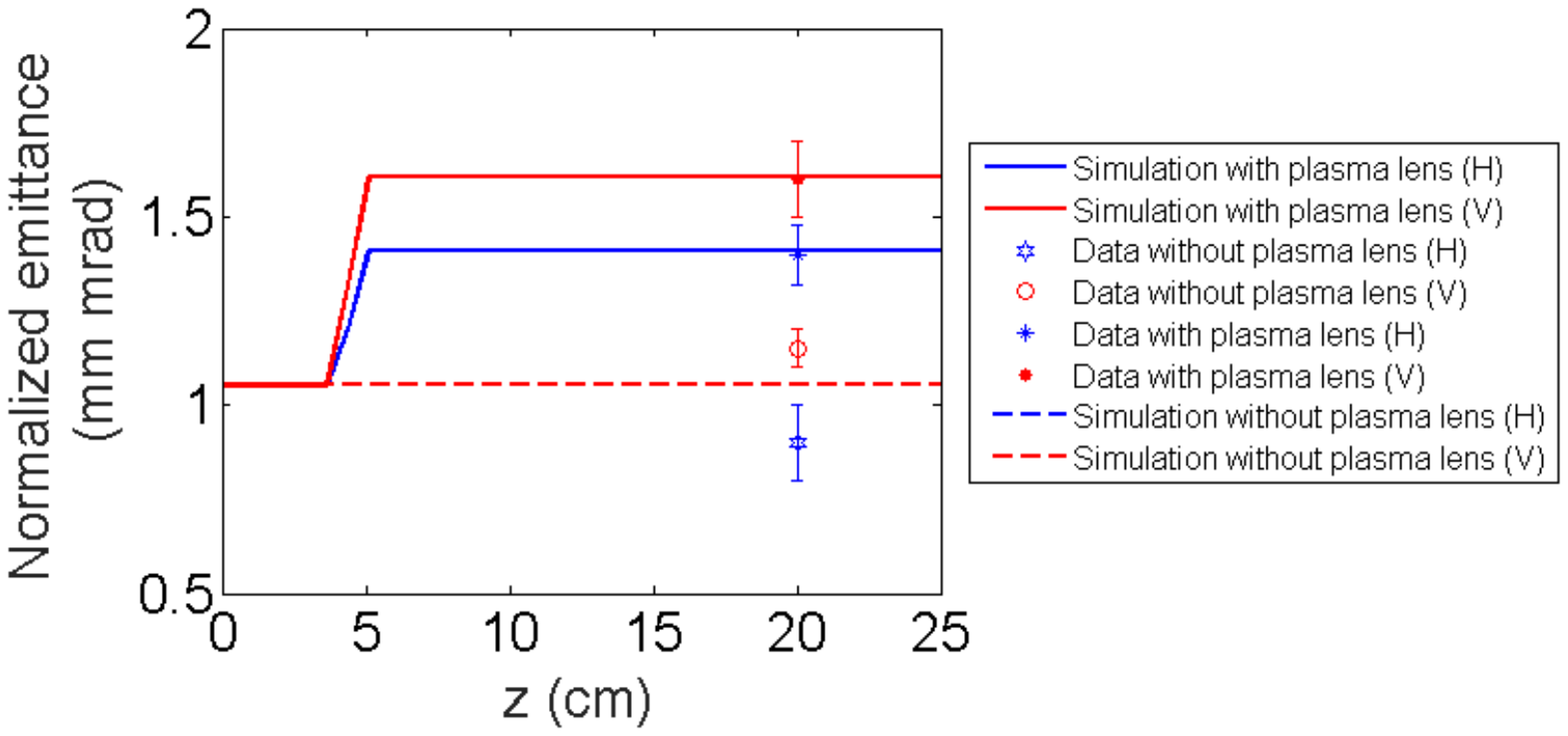}
\caption{Normalized transverse emittance as function of $z$. The numerically computed emittance is plotted with solid lines, and the experimental values are overlapped with stars ($\varepsilon_{nx}$) and circles ($\varepsilon_{ny}$). } 
\label{Fig:comparison}
\end{figure}

Also in this case we have performed simulations to validate our hypothesis and cross-check the measurements. The magnetic profile, retrieved from the 1D simulation, has been included in the simulations performed using GPT~\cite{GPTpaper} and Architect~\cite{marocchino2016efficient}, showing very good agreement with measured data, as shown in Fig.~\ref{Fig:comparison}. An emittance growth between 30 and 40\% has been registered in the plasma, with respect to the case without plasma, probably due to non linearities in the magnetic field, explored by the outer particles of the bunch. 
The horizontal axis, in Fig.~\ref{Fig:comparison} shows a relative longitudinal coordinate along the linac. In this relative coordinate system, the plasma lens is placed at approximately z=5 cm, where the emittance (solid line) starts increasing in the 1 cm long capillary. This is the result of a beam dynamics simulation with the GPT code, assuming in the plasma capillary a radial profile of the azimuthal magnetic field as shown in Fig.~\ref{Fig:IonizDegree}. The dashed line in Fig.~\ref{Fig:comparison} represents the simulated emittance value in case of free space (no plasma in the capillary). Since the the emittance is measured through the conventional quadrupole scan technique about 5 m downstream from the plasma lens and the Twiss parameters are backtracked to the flag at about 20 cm from the plasma, the data of measured emittance (x and y) with (star and dot) and without plasma (hexagon and circle) are both drawn at z=20 cm. 

The experimental results reported in Fig.~\ref{Fig:comparison} are obtained with beam parameters at the plasma entrance typical of a high brightness photo-injector (see Table~\ref{tab:Beam}), in particular characterized by an energy spread much less than 1\%. In case the injected beam in the plasma lens comes from a plasma-accelerating module, with an energy spread of the order of 1\%, the configuration with the plasma lens would be still more advantageous than that with a PMQ triplet because the plasma lens is focusing in both planes, thus the betatron functions stay much smaller in the line and thus the sensitivity to chromatic aberrations is less~\cite{WP5Milestone3}.

\section{Conclusions}
Plasma lenses allow radial focusing with gradient of the order of kT/m and adjustable focal length in case of active configuration and up to MT/m in case of passive configuration. This means that a compact, cm-scale, lens allows for a lower chromaticity system. In particular, progress on active plasma lens experiments at SPARC$\_$LAB has been reported, showing a better control of the emittance growth in the plasma-discharge capillary. We have demonstrated that a shorter capillary contributes to a more uniform distribution of the plasma temperature, resulting in a higher and more constant ionization degree over the capillary radius, due to the higher discharge current, needed to preserve the focal length.

\section*{Acknowledgments}
This work has been partially supported by the EU Commission in the Seventh Framework Program, Grant Agreement 312453-EuCARD-2, the European Union Horizon 2020 research and innovation program under Grant Agreement No. 653782 (EuPRAXIA).

I would like to acknowledge here those colleagues who provided me with material; in particular, in alphabetical order: W. Leemans, J. van Tilborg and BELLA team, E. Adli, C. Lindstr{\o}m and CLEAR team, and J. Osterhoff, J. H. R{\"o}ckemann and FLASHForward collaboration.


\bibliographystyle{elsarticle-num}
\bibliography{biblio}







\end{document}